\documentclass[12pt]{iopart}

\usepackage{iopams}  
\usepackage{graphicx}
\usepackage{epstopdf}
\usepackage{afterpage}
\usepackage{braket}
\usepackage{comment}
\usepackage{ulem}
\usepackage{color}
\usepackage{bm}
\usepackage{setspace}
\usepackage[justification=justified]{caption}
\usepackage{lipsum}

\begin{document}
\
\title[]{Surface plasmon polaritons in thin-film Weyl semimetals}
\author{Tomohiro Tamaya${}^{1*}$, Takeo Kato${}^1$, Kota Tsuchikawa${}^{2,3}$\footnote{Present address: Department of Physics, Tokyo Metropolitan University, Hachioji, 192-0397, Tokyo}, Satoru Konabe${}^4$ and Shiro Kawabata$^3$}
\address{$^1$ Institute for Solid State Physics, University of Tokyo, Kashiwa, 277-8581, Japan}
\address{$^2$ Department of Physics, Tokyo University of Science, Tokyo 162-8601, Japan}
%\address{$^2$ Department of Physics, Tokyo Metropolitan University, Hachioji, 192-0397, Japan}
\address{$^3$ Nanolectronics Research Institute (NeRI), National Institute of Advanced Industrial Science and Technology (AIST), Tsukuba, Ibaraki 305-8568, Japan}
\address{$^4$ Faculty of Bioscience and Applied Chemistry, Hosei University, 3-7-2 Kajino-cho, Koganei, Tokyo 184-8584, Japan}
\ead{tamaya@issp.u-tokyo.ac.jp}
\vspace{10pt}
\begin{indented}
\item[]January 2018
\end{indented}

\begin{spacing}{1.2}

%======================================================
% Title, Author, Affiliation
%======================================================

\begin{abstract}
{We theoretically investigate surface plasmon polaritons propagating in the thin-film Weyl semimetals.
We show how the properties of {surface plasmon polaritons} are affected by hybridization between {plasmons} localized at the two {metal-dielectric} interfaces. 
Generally, this hybridization results in new mixed {plasmon modes}, which are called short-range surface plasmons and long-range surface plasmons, respectively. 
We calculate dispersion curves of these mixed modes for three principle configurations of the axion vector describing axial anomaly in Weyl semimetals. 
We show that the partial lack of the dispersion and the non-reciprocity can be controlled by fine-tuning of the thickness of the Weyl semimetals, the dielectric constants of the outer insulators, and the direction of the axion vector.}
\end{abstract}
%\linenumbers
%======================================================
% Introduction
%======================================================
\section{Introduction}

Surface plasmon polaritons (SPPs) are collective excitations of electrons that propagate along a metal-insulator interface~\cite{Maier,Cai,Sarid00}. 
SPPs can be used to manipulate electromagnetic energy at sub-wavelength scales and provide a route for developing novel devices, such as surface plasmon resonance sensors~\cite{Raether,Homola} and scanning near field optical microscopes~\cite{Novotny,Durig}. 
Recent studies have indicated possibilities for new control methods of SPPs by utilizing two-dimensional Dirac materials~\cite{Liu,Zhu, Lu0,Lu2,Yatooshi,Li00,Cheng00,Fan,Lu3,Cheng01,Cheng02,Cheng03,Li000,Koppens}, topological insulators~\cite{Karch,Poyli,Ou,Yin,Dubrovkin,Guozhi}, or an external static magnetic field~\cite{Liu,Chin,Lin000,Temnov0,Temnov1}. 
In particular, some of these studies point to the chance of developing new optical devices made of materials incorporating the unique properties of SPPs.

Weyl semimetals (WSMs) are potentially such innovative materials~\cite{Armitage,Murakami,Wan,Yang00,Burkov,Xu,Tominaga}.
The WSMs with broken time-reversal symmetry, which have recently been reported for $\rm{YbMnB_{2}}$~\cite{Borisenko} and $\rm{Eu_{2} Ir_{2} O_{7}}$~\cite{Sushkov}, are expected to show unusual optical responses associated with axial anomaly originating from their topological nature~\cite{Zyuzin,Chen0000,Vazifeh,Basar,Pellegrino000,Halterman000}. 
The axial anomaly gives rise to a coupling between material's magnetic and electric properties and leads to chiral magnetic phenomena such as the anomalous Hall effect. 
These phenomena can be expressed in terms of the Maxwell equations in the presence of an axion field~\cite{RLi,Zyuzin000,Baenes16,Wilczek}, and by utilizing them, we expect to find unique SPP properties of WSMs that could be put to use in optical devices.

SPPs propagating along an interface between a bulk WSM and an insulator have been studied theoretically~\cite{Hofmann16,Andolina000}. 
The SPP dispersion at the WSM-insulator interface disappears in an intermediate frequency region and may show non-reciprocity~\cite{Hofmann16}. 
These properties are similar to those in the presence of magnetic fields (surface magnetoplasmons) of up to several tens of tesla ~\cite{Chiu,Hartstein74,Kushwaha01,Wallis74,Brian172,Brian73,Brian74,Kushwaha87a,Kushwaha87b,Kushwaha87c,Kushwaha88a,Halevi88b}. 
Stable and efficient control of magneto-plasmonic SPPs should be possible in WSMs, because the topological nature of these materials fixes the strength of the electro-magnetic coupling originating from chiral magnetic effect, and consequently, can provide a robust platform for experiments and applications. 

WSMs are usually sandwiched between two insulators in actual applications.
In thin-film WSMs, SPPs localized at the two WSM-insulator interfaces hybridize and form new mixed SPP modes, which are called short range surface plasmons (SRSP) and long range surface plasmons (LRSP), respectively~\cite{Fukui,Sarid,Burke,Yang}. 
This indicates that fine tuning of the thickness of the WSM can control SPP dispersion via change of the strength of the hybridization.
In addition, the wavefunctions of the mixed SPP modes can be controlled by changing the dielectric constants of the outer insulators. 
Thus, the thickness of the WSMs and the dielectric constants of the insulators can be regarded as fine-tuning parameters, and hence, revealing their influence on the dispersion of SPPs would provide a useful perspective on how to control them in WSMs.
{We note that quite recently dispersion of the SPP modes has been studied for the case that the axion vector is perpendicular to both a propagation direction and a normal vector of a surface~\cite{Kotov18}.
The thickness dependence and the other configuration of the axion vector have not been investigated yet.}

In this paper, by considering a thin-film WSM sandwiched between two insulators, we theoretically investigate how the properties of SPPs change depending on the thickness of the WSM, the dielectric constants of the outer insulators, and the directions of the axion field. 
Our numerical results indicate that the non-reciprocity with respect to the SPP propagation as well as the disappearance of the LRSP modes can be controlled by tuning the above parameters. 
This conclusion provides a perspective on methods of stably controlling SPPs in WSMs even without an external static magnetic field and pave a way for development of new optical devices such as plasmonic waveguides and optical one-way waveguides.

The organization of this paper is as follows. 
The next section presents the theoretical framework of SPPs in a thin-film WSM sandwiched between outer insulators. 
Since the properties of the SPPs depend on the direction of the axion field, we focus on three principal configurations for the axion vector.
Section~\ref{Results} shows the numerical results for each configuration and discusses their SPP properties.
In section~\ref{Propagation}, we present the characteristic decay lengths for each configuration and refer to their unique characteristics.
Section~\ref{Summary} summarizes the conclusions of this paper.

%======================================================
% Theoretical Framework
%======================================================

\section{Theoretical Framework}
\label{Framework}
\subsection{Maxwell equations with axion modifications}
The unique optical responses in WSMs can be captured by including axion terms in the Lagrangian originating from the topological nature of the materials. 
By considering WSMs with broken time-reversal symmetry, we assume that the axion field $\theta$ only depends on space {(see Supplementary materials)}.
The corresponding Maxwell equations with axion modifications can be derived as\cite{Baenes16,Wilczek}
\begin{eqnarray}
& & \nabla \times \bm{E}=-\frac{\partial \bm{B}}{\partial t}, \label{eq:maxwell1} \\
& & \nabla \cdot \bm{B}=0,  \label{eq:maxwell2} \\
& & \nabla \cdot \bm{D}=\bm{g} \cdot \bm{B}, \label{eq:maxwell3} \\
& & \nabla \times \bm{H}=\frac{\partial \bm{D}}{\partial t}-\bm{g} \times \bm{E}, \label{eq:maxwell4}
\end{eqnarray}
where $\bm{D}$, $\bm{E}$, $\bm{H}$, and $\bm{B}$ are the dielectric displacement, electric field, magnetic field, and magnetic flux density, respectively. 
Here, the vector $\bm{g}=\nabla \theta$ is related to the distance between Weyl nodes {with different chirality} in momentum space, $\bm{b}$: 
\begin{eqnarray} 
\bm{g}=\frac{2 \alpha_0}{\pi}  \sqrt{\frac{\epsilon_{0}}{\mu_{0}}}\bm{b}, 
\end{eqnarray} 
where {$\alpha_0=e^2/\hbar v_{\rm F} \epsilon_{\infty}$} is the fine-structure constant {of the WSM}, $\epsilon_{0}$ is the permittivity of the vacuum, $\mu_{0}$ is the magnetic permeability of the vacuum, {and $v_{\rm F}$ is the Fermi velocity}. 
{Here, we have assumed that the WSM has two Wyle points with different chirality.}
The additional term, $-{\bm g}\times {\bm E}$, expresses the anomalous Hall current, whereas ${\bm g}\cdot {\bm B}$ is the charge induced by its spatial distribution {(note that the anomalous Hall current ${\bm j} = -{\bm g} \times {\bm E}$ and the induced charge $\rho = \bm{g} \cdot \bm{B}$ satisfy the continuity equation $\nabla {\bm j} + \partial \rho/\partial t = 0$ using Eqs.~(\ref{eq:maxwell1}) and (\ref{eq:maxwell2}))}.

\begin{figure}[tb]
\begin{center}
\includegraphics[width=10cm]{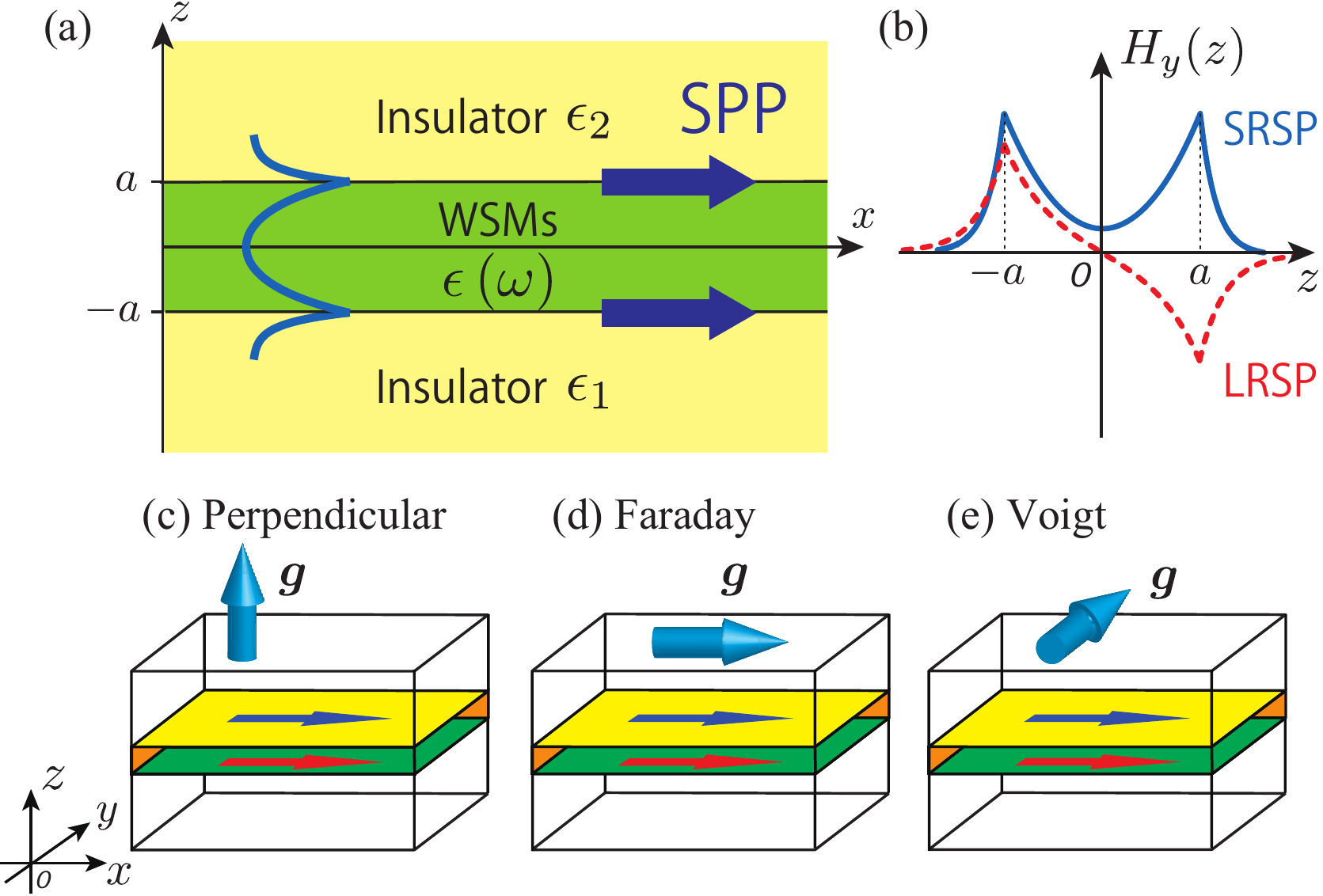}
\caption{(Color online) (a) Setup for studying mixing of two SPP modes at the upper and lower interfaces of WSMs. The $z$-axis is in the stacking direction, while the SPP propagates in the $x$-direction. The thickness of the WSMs is $2a$. The relative dielectric constants of the lower and upper insulators are denoted as $\epsilon_1$ and $\epsilon_2$, respectively, while that of the WSM is $\epsilon(\omega)=\epsilon_\infty(1-\omega_p^2/\omega^2)$. (b) Schematic profiles of $H_y(z)$ for two mixed SPP modes in the limit $q\rightarrow 0$ for $\epsilon_1 = \epsilon_2$. (c)-(e) Schematic diagrams of SPP configurations depending on the axion vector ${\bm g}$: (a) perpendicular configuration (${\bm g}=(0,0,g)$), (b) Faraday configuration (${\bm g}=(g,0,0)$), and (c) Voigt configuration (${\bm g}=(0,g,0)$).}
\label{fig:config}
\end{center}
\end{figure}

We will consider WSMs with broken time-reversal symmetry $(\bm{g}\neq {\bm 0})$ sandwiched between two insulators with relative dielectric constants $\epsilon_i$ ($i=1,2$), as shown in Fig.~\ref{fig:config}~(a). 
The $z$-axis is in the stacking direction, and the region of the WSM layer is $-a < z < a$, where $2a$ is the thickness of the WSM. 
We will assume that SPPs propagate in the $x$-direction. 
Accordingly, the electric and magnetic fields of the SPPs can be written as follows:
\begin{eqnarray}
     & & ({\bm E}({\bm r},t))_\alpha = E_{\alpha}(z) e^{iqx-i\omega t}, \\
     & & ({\bm H}({\bm r},t))_\alpha = H_{\alpha}(z) e^{iqx-i\omega t}, 
\end{eqnarray}
where $q$ and $\omega$ are the wavenumber and frequency of the SPPs, respectively, and $\alpha = x,y,z$. 
The {wavefunctions} of the electromagnetic field in the two outer insulators are $X_{\alpha}(z) = X_{\alpha,1} e^{\kappa_1 z}$ ($z<-a$) and $X_{\alpha,2} e^{-\kappa_2 z}$ ($z>a$), where $X=E$ or $H$, $E_{\alpha,i}$ and $H_{\alpha,i}$ ($i=1,2$) are complex constants, $\kappa_i =\sqrt{q^{2}-\epsilon_i (\omega/c)^{2}}$ ($i=1,2$), and $c=(\epsilon_0 \mu_0)^{-1/2}$ is the velocity of light in a vacuum. 
Since the electromagnetic field of SPP is localized at and decays exponentially away from the interface, the $\kappa_i$ must be positive real numbers.

Let us briefly summarize the properties of the SPP dispersion in the absence of axion fields ($\bm{g}= {\bm 0}$), which corresponds to a normal metal sandwiched between two insulators.
SPPs are always described with the transverse magnetic (TM) mode, because the transverse electric (TE) mode cannot support confined SPPs, except when either the magnetic permeability or dielectric constant is negative \cite{Maier,Cai,Sarid00}. 
As the thickness of the WSM layer decreases, the two SPP modes at the upper and lower interfaces start to hybridize and change into two mixed SPP modes. 
The low- and high-frequency mixed modes are called the short range surface plasmons (SRSP) and the long range surface plasmons (LRSP), respectively, because the penetration depth $\kappa_i^{-1}$ ($i=1,2$) in the insulators are longer for LRSPs than for SRSPs. 
For $\epsilon_1=\epsilon_2$, the SRSPs and LRSPs have symmetric and antisymmetric profiles of the transverse magnetic field $H_y(z)$ for $q\rightarrow 0$ (see Fig.~\ref{fig:config}~(b)), respectively. 
For $q\rightarrow \infty$, these two modes are localized at one of the two interfaces, depending on the magnitude correlation between $\epsilon_1$ and $\epsilon_2$.
{We note that the thickness of WSMs considered here is of order of sub-micrometer, for which effect of a finite-size gap for Weyl cone is negligible (the finite-size gap is considered to be caused by the atomic scale thickness of WSMs, which demands more complex and serious atomistic/ab initio theories).}

In what follows, we consider three configurations for the axion vector ${\bm g}$ (see Fig.~\ref{fig:config}~(c)-(e)): (i) the perpendicular configuration ($\bm{g}=(0,0,g)$), (ii) the Faraday configuration ($\bm{g}=(g,0,0)$), and (iii) the Voigt configuration ($\bm{g}=(0,g,0)$).
Here, for the configurations, we have used the terminology in the study of the surface magneto plasmons\cite{Chiu,Hartstein74,Kushwaha01,Wallis74,Brian172,Brian73,Brian74,Kushwaha87a,Kushwaha87b,Kushwaha87c,Kushwaha88a,Halevi88b,Zhukov00}, since the chiral magnetic effect is similar to that of an external magnetic field.
For simplicity, the dielectric function of the WSM layer is taken as $\epsilon(\omega)=\epsilon_\infty (1-\omega_p^{2}/\omega^{2})$ following Ref.~\cite{Hofmann16}, where $\epsilon_\infty$ is the relative dielectric constant in the limit $\omega \rightarrow \infty$, and $\omega_p$ is the plasma frequency of the WSM layer. 
{We note that this simplified dielectric function can give reasonable results on SPPs because almost the same results are obtained using more realistic dielectric function based on the Kubo formula, as long as considering below the plasma frequency\cite{Kotov16}.}
We will only focus on the low-frequency region $\omega < \omega_p$, for which the dielectric function $\epsilon(\omega)$ is always negative.
Detail information on the following calculations is summarized in the Supplementary materials.

\subsection{Perpendicular configuration}
For the perpendicular configuration ($\bm{g}=(0,0,g)$), the axion term mixes the TE and TM modes\cite{Hofmann16}. 
The {wavefunction} of the electromagnetic field in the WSM layer is a linear combination of four elementary solutions of the modified Maxwell equations,
\begin{eqnarray}
& & X_{\alpha}(z) = X_{\alpha}^{++} e^{-\kappa_+ z} + X_{\alpha}^{+-} e^{-\kappa_- z} \nonumber \\
& & \hspace{10mm} + X_{\alpha}^{-+} e^{\kappa_+ z} + X_{\alpha}^{--} e^{\kappa_- z}, \quad (X=H \, {\rm or} \, E),
\label{eq:perpHy}
\end{eqnarray}
for $|z|<a$, where $\kappa_\pm$ are real numbers defined as
\begin{eqnarray}
& & \kappa_{\pm}^2=q^2-\epsilon(\omega) \omega^2/c^2
\nonumber \\
& & \hspace{10mm} \pm g\sqrt{\frac{\mu_{0}}{\epsilon_0 \epsilon(\omega)}(\epsilon(\omega)\omega^2/c^2 -q^2)}.
\label{eq:perpkappa}
\end{eqnarray}
Upon eliminating $E_x$, $E_y$, $D_z$, $B_x$, $B_y$, and $B_z$ from the modified Maxwell equations by using the continuity conditions at the interfaces, we obtain the linear equation,
\begin{eqnarray}
\hat{U}(q,\omega) \left( \begin{array}{c} 
H_{y}^{++} \\ H_{y}^{+-} \\ H_{y}^{-+}  \\ H_{y}^{--}
\end{array} \right) = 0,
\label{eq:defU}
\end{eqnarray}
where $\hat{U}(q,\omega)$ is a $4\times 4$ matrix. 
The dispersion relation $\omega = \omega(q)$ is determined by the condition ${\rm det} \, \hat{U}(q,\omega)= 0$, for which nontrivial solutions of (\ref{eq:defU}) exist. 
After obtaining the dispersion, we have to check whether the condition $\kappa_-^2 > 0$ is satisfied or not, since the SPP modes should be localized at the interfaces. 
For the perpendicular configuration, the SPP dispersion has the reciprocity relation, i.e., $\omega(-q) = \omega(q)$.

\subsection{Faraday configuration}
The axion term mixes the TE and TM modes as well for the Faraday configuration ($\bm{g}=(0,0,g)$). 
The wavefunction of the SPP modes in the WSM layer is given by Eq.~(\ref{eq:perpHy}), where $\kappa_\pm$ are replaced with
\begin{eqnarray}
& & \kappa_{\pm}^2=q^2-\epsilon(\omega) \omega^2/c^2 +\frac{g^2\mu_{0}}{2\epsilon_0 \epsilon(\omega)} \pm g \sqrt{K}, \\
& & K = \frac{\mu_{0}}{\epsilon_0 \epsilon(\omega)} \left(q^2+ \frac{g^2 \mu_0}{4\epsilon_0 \epsilon(\omega)} \right)
\label{eq:kappaDDef}
\end{eqnarray}
By eliminating variables, we obtain the linear equation in the form of Eq.~(\ref{eq:defU}). 
The dispersion relation is determined by ${\rm det} \, \hat{U}(q,\omega)= 0$. 
For the Faraday configuration, $\kappa_\pm$ become complex when $K<0$ for ${\bm g} \ne {\bm 0}$. 
The corresponding solutions, called complex solutions~\cite{Kushwaha01}, indicate that $H_y(z)$ spatially oscillates and exponentially decays in the $z$-direction. 
The SPP dispersion has the reciprocity relation, $\omega(-q) = \omega(q)$, for the Faraday configuration.

\subsection{Voigt configuration}

For the Voigt configuration, the axion field does not mix the TE and TM modes. 
Therefore, the {wavefunction} of the electromagnetic field in the WSM is a linear combination of two elementary solutions of the modified Maxwell equations:
\begin{eqnarray}
X_{\alpha}(z) = X_{\alpha}^+ e^{-\kappa z} + X_{\alpha}^- e^{\kappa z}, \quad (X=H \, {\rm or} \, E),
\label{eq:VoigtHy}
\end{eqnarray}
for $|z|<a$, where $\kappa$ is a real number defined as
\begin{eqnarray}
\kappa^2 = q^2-\frac{\mu_0}{\epsilon_{0}\epsilon(\omega)}\left((\omega \epsilon_{0}\epsilon(\omega))^2-g^2\right).
\end{eqnarray}
After eliminating $E_x$, $D_z$, and $B_y$ using the continuity condition, we obtain the linear equation, 
\begin{eqnarray}
\hat{V} \left( \begin{array}{c} 
H_y^+ \\ H_y^- 
\end{array} \right) = 0,
\label{eq:defV}
\end{eqnarray}
where $\hat{V}$ is a $2\times 2$ matrix. 
The SPP dispersion is determined by the equation ${\rm det} \, \hat{V}(q,\omega)= 0$. 
In the Voigt configuration, the SPP dispersion generally has the non-reciprocity relation, $\omega(-q) \ne \omega(q)$, because the determinant equation of the SPP is not invariant under the reversal of the propagation direction $q \rightarrow -q$.

\section{Results and Discussion}
\label{Results}
This section presents numerical results on the SPP dispersion for the three configurations of ${\bm g}$. We take the amplitude of the axion vector to be $g=4 {\epsilon_0} \omega_{P}$.
Furthermore, we take the relative dielectric constant of the WSMs for $\omega \to \infty$ to be $\epsilon_{\infty}=13$ as measured in ${\rm Eu}_2{\rm Ir}_2{\rm O}_7$~\cite{Hofmann16}.

\begin{figure}[tb]
\begin{center}
\includegraphics[width=14cm]{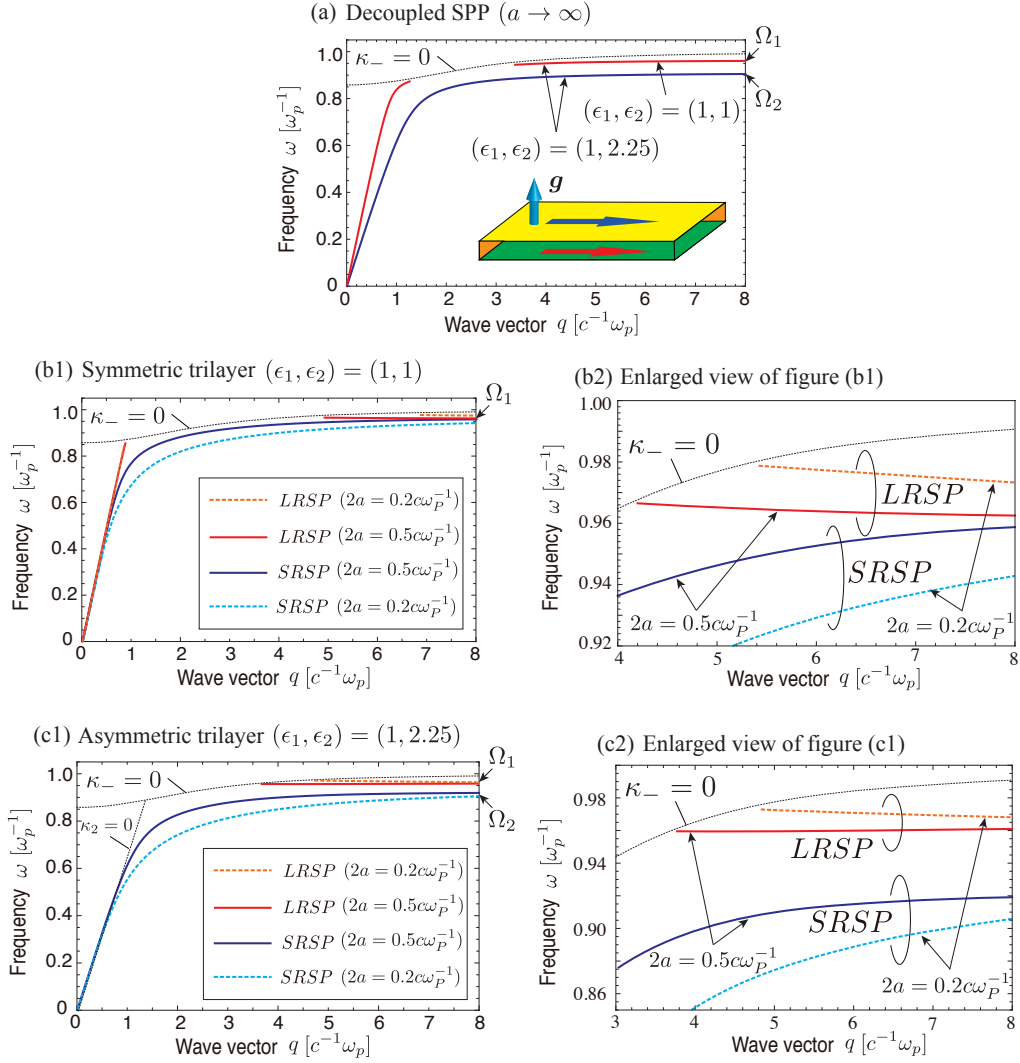}
\caption{(Color online) {SPP dispersion for the perpendicular configuration. (a) SPP dispersion in the limit $a \rightarrow \infty$. Blue and red lines indicate dispersion curves tangent to air $(\epsilon_{i}=1)$ and glass $(\epsilon_{i}=2.25)$, respectively. (b1) Dispersion curves of the mixed SPP modes, SRSP (blue and light blue lines) and LRSP (orange and red lines) generated from mutual interference of wavefunctions in the cases of $2a=0.5 c\omega_{\rm P}^{-1}$ (the solid lines) and $0.2 c\omega_{\rm P}^{-1}$ (the dashed lines) for a symmetric trilayer system $(\epsilon_1,\epsilon_2) = (1,1)$. (c1) Dispersion curves of mixed SPP modes for an antisymmetric trilayer system $(\epsilon_1,\epsilon_2) = (1,2.25)$. (b2), (c2) Enlarged view of figures (b1) and (c1), respectively, in the range of $4 c \omega_{P}^{-1} \leq q \leq 8c \omega_{P}^{-1}$ and $3 c \omega_{P}^{-1}\leq q \leq 8 c \omega_{P}^{-1}$. The dotted lines indicate the condition $\kappa_- = 0$, and $\Omega_1$ and $\Omega_2$ are determined by $\epsilon(\Omega_1)=-1$ and $\epsilon(\Omega_2)=-2.25$, respectively.}}
\label{fig:perpendicular}
\end{center}
\end{figure}

\subsection{Perpendicular configuration}

First, let us consider the limit $a\rightarrow \infty$. 
Since there is no mutual interference between SPPs at the two interfaces, the dispersion is individually determined at the upper and lower interfaces, respectively. 
Figure~\ref{fig:perpendicular}~(a) plots the individual SPP dispersion at the WSM-insulator interface. 
The upper and lower curves correspond to SPPs propagating along the interface tangent to insulators with relative dielectric constants of $\epsilon=1$ and 2.25. 
When the dielectric constants of the outer insulators are $(\epsilon_{1},\epsilon_{2}) =(1,1)$ (only vacuum), the dispersion curves of the SPPs are degenerate, both following the upper curve. 
For $(\epsilon_{1},\epsilon_{2}) =(1,2.25)$ (vacuum and glass), the dispersion curves of the SPPs differ at the two interfaces and correspond to the upper and lower curves. 
In the long-wavelength limit ($q\rightarrow 0$), the dispersions become linear with inclinations $c/\sqrt{\epsilon_{1}}$ and $c/\sqrt{\epsilon_{2}}$. In the short-wavelength limit ($q\rightarrow \infty$), they approach constant frequencies $\Omega_{1}$ and $\Omega_{2}$. 
In the case of $(\epsilon_1,\epsilon_2)=(1,2.25)$, $\Omega_{1}$ and $\Omega_{2}$ are respectively determined by the conditions $\epsilon(\Omega_1) = -1$ and $\epsilon(\Omega_2) = -2.25$. 
It is remarkable that part of the SPP dispersion curve may disappear at the intermediate wavenumber for the case of interfaces tangent to a vacuum, because they encroach on the condition $\kappa_{-}= 0$ and come to have bulk-propagating modes in the $z$-direction. 
Moreover, we can see that the dispersion curve is shifted down as the dielectric constants of the insulators increase and the interval of the disappearance of SPPs becomes narrower.
The lack of the SPP dispersion is characteristic of axion electromagnetic dynamics, and it originates from the anomalous Hall current induced by the axion field\cite{Hofmann16}.

\begin{figure}[tb]
\begin{center}
\includegraphics[width=14cm]{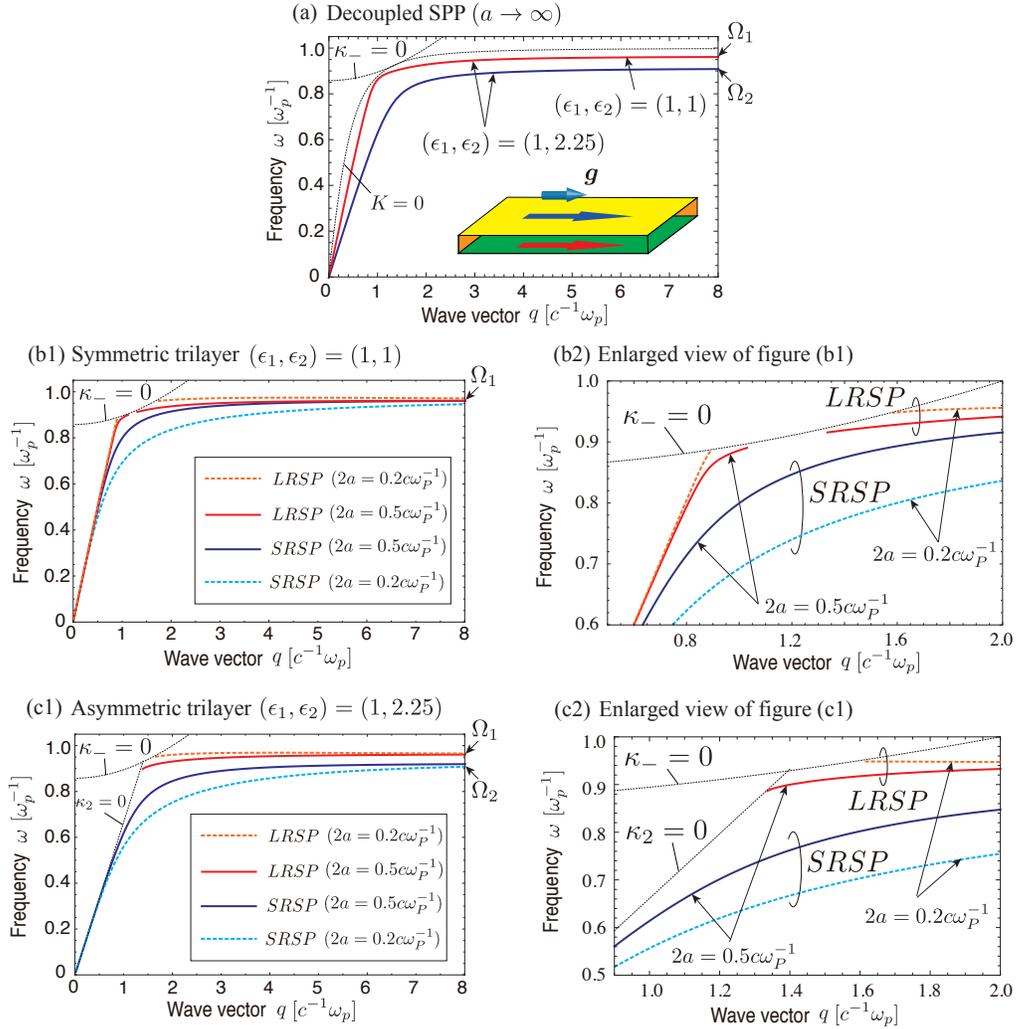}
\caption{(Color online) {SPP dispersion for the Faraday configuration. (a) SPP dispersion in the limit $a \rightarrow \infty$. Blue and red lines indicate dispersion curves tangent to air $(\epsilon_{i}=1)$ and glass $(\epsilon_{i}=2.25)$, respectively. The lower dotted line in Fig. (a) indicates the condition $K=0$, where $K = ({\mu_{0}}/{\epsilon_0 \epsilon(\omega)}) \left(q^2+{g^2 \mu_0}/{4\epsilon_{0}\epsilon(\omega)}\right)$. (b1) Dispersion curves of the mixed SPP modes, SRSP (blue and light blue lines) and LRSP (orange and red lines) generated from mutual interference of wavefunctions in the cases of $2a=0.5 c\omega_{\rm P}^{-1}$ (the solid lines) and $0.2 c\omega_{\rm P}^{-1}$ (the dashed lines) for a symmetric trilayer system $(\epsilon_1,\epsilon_2) = (1,1)$. (b2) Dispersion curves of mixed SPP modes for an antisymmetric trilayer system $(\epsilon_1,\epsilon_2) = (1,2.25)$. (b2), (c2) Enlarged view of figures (b1) and (c1), respectively, in the range of $0.6 c \omega_{P}^{-1} \leq q \leq 2.0c \omega_{P}^{-1}$ and $0.9 c \omega_{P}^{-1}\leq q \leq 2.0 c \omega_{P}^{-1}$. The dotted lines indicate the condition $\kappa_- = 0$, and $\Omega_1$ and $\Omega_2$ are determined by $\epsilon(\Omega_1)=-1$ and $\epsilon(\Omega_2)=-2.25$, respectively.}}
\label{fig:Faraday}
\end{center}
\end{figure}

Figures~\ref{fig:perpendicular}~(b) and (c) show SPP dispersion curves for thin WSM layers with $2a=0.2 c\omega_p^{-1}$ and $0.5 c\omega_p^{-1}$. 
These curves result from mutual interference of the SPPs at the upper and lower interfaces, and they correspond to the mixed modes called LRSP and SRSP (see Sec.~\ref{Framework}). 
For the symmetric trilayers $(\epsilon_1,\epsilon_2)=(1,1)$ (Fig.~\ref{fig:perpendicular}~(b)), we can see that the degenerate SPP modes (the upper curve in Fig.~\ref{fig:perpendicular}~(a)) split into the LRSP and SRSP modes. 
The hybridization between SPPs localized at the two interfaces becomes more significant when the thickness of the WSMs decreases, and the interval in which the SPP dispersion disappears becomes more (less) significant for the LRSP (SRSP) mode. 
Since the boundary $\kappa_{-} = 0$ is independent of $a$, the interval in which the SPP dispersion disappears becomes more (less) significant for the LRSP (SRSP) mode as $a$ decreases.
This can be confirmed in Fig.~\ref{fig:perpendicular}~(b), where the interval in which the LRSP mode disappears becomes broader with decreasing WSM thickness, while the SRSP mode exists in the whole region. 
We should note that in the short-wavelength limit ($q \to \infty $), the dispersion curves of the LRSP and SRSP modes converge to the same frequency determined by the condition $\epsilon(\Omega_1) = -1$.

Mutual interference of the SPP modes also occurs in the asymmetric trilayers $(\epsilon_1,\epsilon_2)=(1,2.25)$ (Fig.~\ref{fig:perpendicular}~(c)). 
As the thickness of WSMs decreases, mode repulsion between the mixed SPPs becomes significant because hybridization of SPPs increases.
In the short-wavelength limit ($q \to \infty$), the dispersion curves of the LRSP and SRSP converge to different frequencies determined, respectively, by the conditions $\epsilon(\Omega_1) = -1$ and $\epsilon(\Omega_2) = -2.25$.
This is because for $q \to \infty$ the wavefunctions of the LRSP and SRSP modes are localized at the lower and upper interfaces tangent to the different insulators, respectively.
In contrast to the symmetric case, the LRSP mode disappears in the long-wavelength region ($q \to 0$), because it is located between the two-light cones, $\omega=q/\sqrt{\epsilon_{1}}$ and $\omega=q/\sqrt{\epsilon_{2}}$, and has bulk-propagating modes in one of the two insulators. 

\subsection{Faraday configuration}

Figure~\ref{fig:Faraday}~(a) shows the SPP dispersions in the Faraday configuration in the limit $a\rightarrow \infty$, where there is no mutual interference between SPPs at the two interfaces. 
The upper and lower curves correspond to SPPs at the interface tangent to insulators with dielectric constants of $\epsilon=1$ and 2.25. 
For $(\epsilon_{1},\epsilon_{2}) =(1,1)$ (only vacuum), the dispersion curves of the SPPs are degenerate, both following the upper curve. 
For $(\epsilon_{1},\epsilon_{2}) =(1,2.25)$ (vacuum and glass), the dispersion curves of the SPPs differ at the two interfaces and correspond to the upper and lower curves.
The features shown in this figure are similar to those of the perpendicular configuration; the SPPs have linear dispersion in the long-wavelength limit ($q \to 0$), while they have constant frequencies in short-wavelength limit ($q \to \infty$). 
The figure also has a plot for the condition $K=0$ (lower thin dotted line), where $K$ is defined as Eq. (12). 
Below this line, the SPP solution has complex decay constants, $\kappa_{\pm}$, which physically mean oscillating decay of the wavefunctions along the $z$-axis inside the WSM. 

Figures~\ref{fig:Faraday}~(b) and (c) show the SPP dispersion for the symmetric trilayers of $(\epsilon_{1},\epsilon_{2}) =(1,1)$ and the asymmetric trilayers of $(\epsilon_{1},\epsilon_{2}) =(1,2.25)$. 
The parameters utilized to make these figures are the same as those of the perpendicular configuration. 
Similar to the results for the perpendicular configuration, these figures clearly show hybridization of the SPP modes. 
For the symmetric trilayers (Fig.~\ref{fig:Faraday}~(b)), the hybridization becomes more significant as the thickness of the WSMs decreases, and the interval in which the SPP dispersion disappears becomes more (less) significant for the LRSP (SRSP) mode. 
For the asymmetric trilayers (Fig.~\ref{fig:Faraday}~(c)), the mode repulsion becomes more significant with decreasing the WSM thickness, and the LRSP mode disappears in the long-wavelength limit ($q \to 0$).
We should note that these modes have complex $\kappa_{\pm}$ when the dispersion curve is below the boundary $K=0$(see Fig. 3(a)), which indicates oscillating decay of the wavefunctions in the $z$-direction. 
It is remarkable that the inverval in which the SPP dispersion disappears is narrower than that of the perpendicular configuration.

\subsection{Voigt configuration}

\begin{figure}[tb]
\begin{center}
\includegraphics[width=14cm]{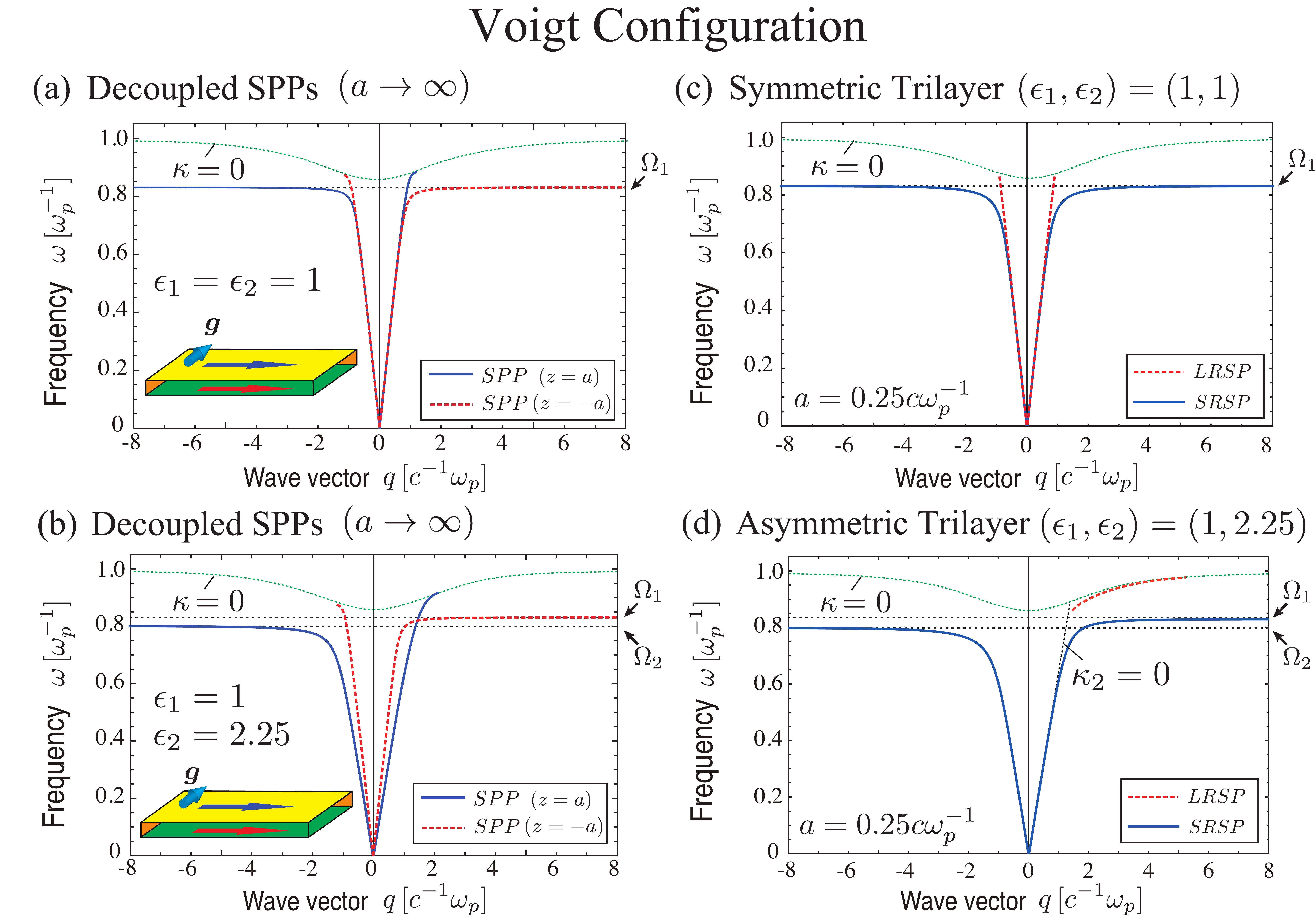}
\caption{(Color online) SPP dispersion for the Voigt configuration. 
Dispersions in the limit $a \rightarrow \infty$ are shown for (a) $(\epsilon_1,\epsilon_2) = (1,1)$ and (b) $(\epsilon_1,\epsilon_2) = (1,2.25)$. The blue solid and red dashed lines indicate the SPP modes localized around $z=a$ and $z=-a$, respectively. The SPP dispersion in a thin WSM layer ($2a=0.5c \omega_{\rm P}^{-1}$), for which the upper and lower boundary wavefunctions mutually interfere, is shown for (c) $(\epsilon_1,\epsilon_2) = (1,1)$ and (d) $(\epsilon_1,\epsilon_2) = (1,2.25)$. The blue solid and red dashed lines indicate the SRSP and LRSP modes, respectively. The upper thin dotted lines show the condition $\kappa = 0$, while the straight dotted line in (d) shows the condition $\kappa_2=0$.}
\label{fig:Voigt}
\end{center}
\end{figure}

In Fig.~\ref{fig:Voigt}~(a) and (b), we show the SPP dispersion in the Voigt configuration in the limit $a \rightarrow \infty$ for the symmetric trilayers $(\epsilon_{1},\epsilon_{2}) =(1,1)$ and the asymmetric trilayers $(\epsilon_{1},\epsilon_{2}) =(1,2.25)$, respectively.
Here, the solid and dashed lines indicate SPPs localized at $z=a$ and $z=-a$ interfaces, respectively. 
For both cases, the SPP dispersion at each interface clearly indicates the non-reciprocity, $\omega(-q)\neq \omega(q)$. 
In the long-wavelength limit ($q \to 0$), the SPP dispersion at the interface $z=\mp a$ becomes linear with inclination $ c/\sqrt{\epsilon_i} (i=1,2)$, whereas in the short-wavelength limit ($q \to \infty$), it approaches a constant frequency $\Omega_{i}$ ($i=1,2$). 
A simple calculation yields $\Omega_{i}$ in the following form {(see Supplementary materials)}:
\begin{eqnarray}
\Omega_i =  \frac{-g+\sqrt{g^2+4\epsilon_\infty \epsilon_{0}^2\omega^2_p(\epsilon_\infty+\epsilon_i)}}{2\epsilon_{0}(\epsilon_\infty+\epsilon_i)}, \quad (i=1,2).
\label{eq:Omegapm}
\end{eqnarray}
In contrast to the perpendicular and Faraday configurations, $\Omega_{i}$ clearly depends on the strength of the axion vector $\bm{g}$. We can also see that the SPP dispersion at $z=-a$ exists in the whole regions for $q>0$, while it disappears for $q<0$ at high frequencies because of the condition $\kappa^2 > 0$ (in the figure, the boundary line $\kappa=0$ is shown by the thin dotted lines). 
For the symmetric trilayers (Fig.~\ref{fig:Voigt}~(a)), the SPP dispersion at $z=-a$ is just a reversal of that at $z=a$, because only the direction of the surface is opposite at two interfaces with respect to the SPP propagation direction.
In contrast, the asymmetric trilayers do not have this property because the dielectric constants of the two outer insulators differ. 

Figures~\ref{fig:Voigt}~(c) and (d) show the SPP dispersion in the presence of the mutual interference between SPPs for the symmetric trilayers $(\epsilon_{1},\epsilon_{2}) =(1,1)$ and the asymmetric trilayers $(\epsilon_{1},\epsilon_{2}) =(1,2.25)$, respectively.
The dashed and solid lines indicate the LRSP and SRSP modes in the case of $2a=0.5c \omega_{\rm P}^{-1}$. 
For the symmetric trilayers, we can see that the hybridization between SPPs recovers the reciprocity in their dispersion. 
The reciprocity can be confirmed analytically by supposing $\epsilon_{1}=\epsilon_{2}$ in the determinant equation (${\rm det} \, \hat{V}=0$) for the Voigt configuration (see the Supplementary materials), which can be expressed as follows:
\begin{eqnarray}
    & & e^{-4\kappa a}\left[(\kappa_1(\omega^{2} \epsilon(\omega)^2+g^2)- \epsilon(\omega) \epsilon_{1}\omega^{2}\kappa)^{2}-\epsilon_{1}^{2}q^2\omega^2g^2\right] \nonumber \\
    &-&\left[(\kappa_1(\omega^{2} \epsilon(\omega)^2+g^2)+\epsilon(\omega) \epsilon_{1}\omega^{2}\kappa)^{2}-\epsilon_{1}^{2}q^2\omega^2g^2\right] = 0. \nonumber 
\end{eqnarray}
This equation includes $q$ in quadratic form and is invariant under the transformation, $q \to -q$. 
The hybridization of the SPPs erases the non-reciprocity, because the mixing of the two SPPs at the two interfaces, whose wavefunctions transform into each other by reversing the propagation direction (see Fig.~\ref{fig:Voigt}~(a)), completely cancels the non-reciprocity. 
This consideration suggests that the non-reciprocity can remain for the asymmetric trilayer system.
Figure~\ref{fig:Voigt}~(d) confirms this conjecture, wherein the dispersion of the mixed SPP modes keeps the non-reciprocity.
The non-reciprocity can also be confirmed directly from the determinant equation ${\rm det} \, \hat{V}=0$ for the Voigt configuration (see the Supplementary materials). 
Figure~\ref{fig:Voigt}~(d) also indicates that the SRSP mode exists in the whole $q$ space, while the LRSP mode exists in a limited region and is only allowed for propagation in the positive $x$-direction. 
We should note that the strength of the axion vector $\bm{g}$ of the WSMs can be experimentally determined by measuring $\Omega_1$ and $\Omega_2$ (see Fig.~\ref{fig:Voigt}~(d)). 

Finally, we should note that the Voigt configuration can be changed into the Faraday configuration in experiments simply by rotating either the incident electromagnetic waves or the materials. 
This opens up the possibility for simple and flexible control of the stability and non-reciprocity of SPPs.

\section{Characteristic decay length}
\label{Propagation}

\begin{figure}[tb]
\begin{center}
\includegraphics[width=7cm]{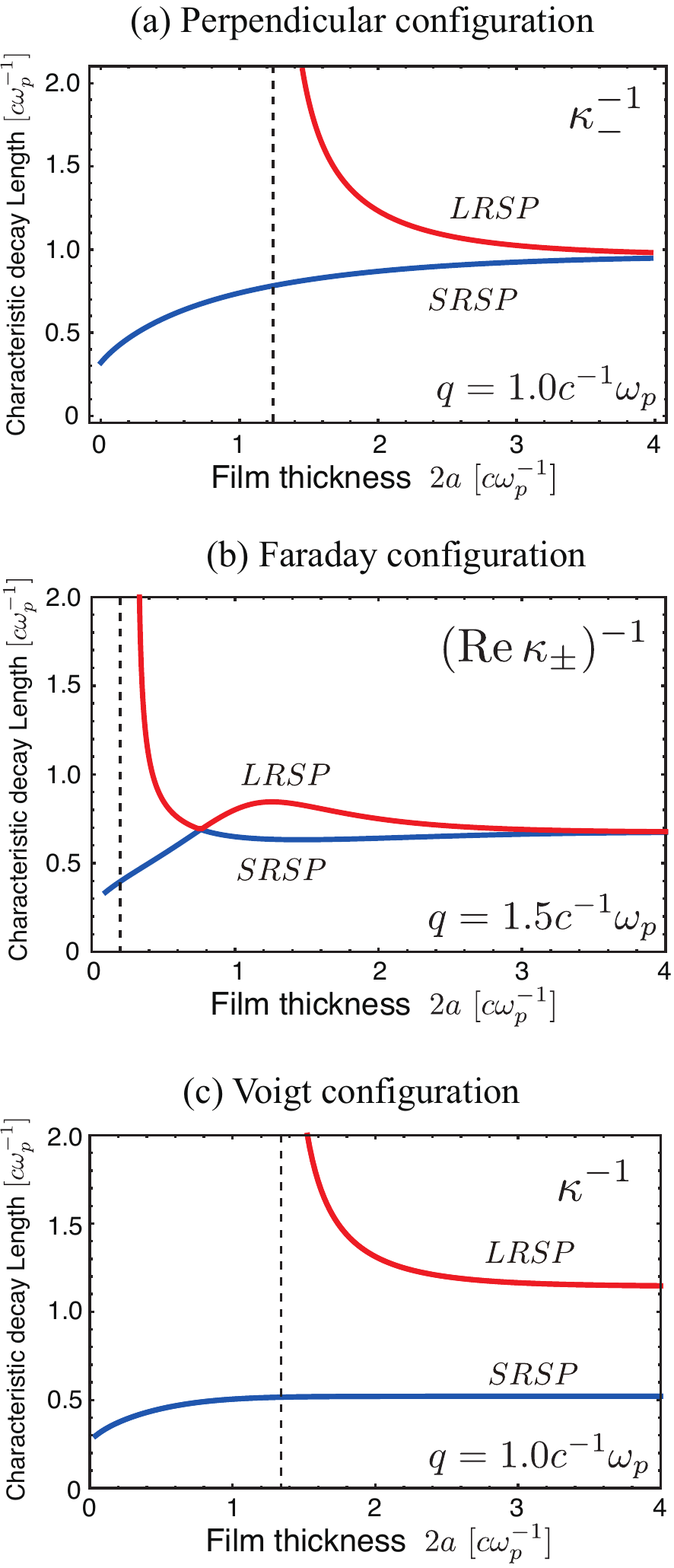}
\caption{(Color online) Characteristic decay length as a function of WSM thickness for (a) the perpendicular configuration, (b) Faraday configuration, and (c) Voigt configuration. The dielectric constants of the insulators are $\epsilon_1=\epsilon_2=1$. The wavenumber of SPPs is $q=1.0c^{-1}\omega_p$ for the perpendicular and Voigt configurations, while it is $q=1.5c^{-1}\omega_p$ for the Faraday configuration.}
\label{fig:depth}
\end{center}
\end{figure}

We first consider the characteristic decay length of the SPPs in WSMs in the stacking direction for the perpendicular configuration.
We should note that the wavefunction given in Eq.~(\ref{eq:perpHy}) includes the two length scales, $\kappa_+^{-1}$ and $\kappa_-^{-1}$.
For this wavefunction, it is difficult to define the penetration depth, which is measurable in experiments.
Here, we focus on the longer length scale $\kappa_-^{-1}$ {(see Eq.~(\ref{eq:perpkappa}))}, since the thickness dependence of $\kappa_+^{-1}$ is weak.
Figure~\ref{fig:depth}~(a) shows $\kappa_{-}^{-1}$ as a function of the WSM thickness $2a$ for $q=c^{-1}\omega_p$ and $\epsilon_1=\epsilon_2=1$.
With decreasing thickness of WSM, the mutual interference at the upper and lower interfaces becomes stronger, and consequently, the characteristic decay lengths of SRSP and LRSP decrease and increase, respectively. 
This result indicates that the hybridization of the SPPs makes the SRSP more stable (more localized at the interface), while it makes the LRSP unstable (less localized at the interface).
The divergence of the characteristic decay length at a critical value of $a$ indicates disappearance of LRSP mode.

Next, we consider the characteristic decay length of the SPPs for the Faraday configuration.
For $q=1.5c^{-1}\omega_p$ and $\epsilon_1=\epsilon_2=1$, $\kappa_\pm$ are always complex numbers and are complex conjugates of each other. 
The characteristic decay length is thus defined as ${\rm Re} [\kappa_{\pm}]^{-1}$. 
Figure~\ref{fig:depth}~(b) shows ${\rm Re} [\kappa_{\pm}]^{-1}$ as a function of the WSM thickness.
In this figure, we can see that the SRSP and LRSP modes cross around $2a=2a_{\rm c}\simeq 0.75c\omega_p^{-1}$, which indicates a non-monotonic properties for splitting between the SRSP and LRSP modes. 
This result enables us to conceive that the symmetry of $H_y(z)$ may change at this crossing point. 
From numerical results, we could identify that SRSP(LRSP) show almost symmetric(antisymmetric) profiles for $H_y(z)$ when $a>a_{\rm c}$ as expected for the usual symmetric trilayers.
The SRSP and LRSP modes, however, exchange their symmetry for $a<a_{\rm c}$; the SRSP(LRSP) show almost antisymmetric(symmetric) profiles for $H_y(z)$. 
In the Faraday configuration, the anomalous Hall current flows mainly along the $y$-axis and has different signs at the two interfaces. 
We suppose that the change of the symmetries at $a=a_{\rm{c}}$ is induced by interference between the spatial profile of the anomalous Hall current and that of the magnetic field $H_y$. 
The existence of a crossing point is a unique property of the Faraday configuration, and a similar effect may be expected for the SPPs of the ferromagnetic materials as well as in the presence of an external static magnetic field {,where the external magnetic and ferromagnetic field have the same direction to that of the axion vector}. 
The divergence of ${\rm Re} [\kappa_{\pm}]^{-1}$ indicates disappearance of the LRSP mode.

For the Voigt configuration, the penetration length is well defined as $\kappa^{-1}$ (see Eq.~(\ref{eq:VoigtHy})).
Figure~\ref{fig:depth}~(c) shows $\kappa^{-1}$ as a function of the WSM thickness.
Because of the non-reciprocity at the upper and lower boundary SPP modes, the characteristic decay lengths of the SRSP and LRSP modes for $a\rightarrow \infty$ becomes different even in the absence of the hybridization. 
With decreasing thickness of the WSM layer, the hybridization becomes significant and the characteristic decay length of LRSP (SRSP) increases (decreases). 
The different behaviors of the the characteristic decay lengths in these configurations would provide new methods for controlling SPPs in WSMs.

%======================================================
% Summary
%======================================================

\section{Summary}
\label{Summary}

We theoretically investigated the properties of SPPs in thin-film WSMs sandwiched between outer insulators.
We considered three configurations of the axion vector ${\bm g}$, i.e., the perpendicular, Faraday, and Voigt configurations.
When the thickness of the WSM film is sufficiently small, SPPs localized at upper and lower boundaries start to hybridize and form mixed SPP modes. 
The mutual interference of the wavefunctions induces a split of the degenerate SPP modes for the symmetric trilayers, while it induces mode repulsion of the two non-degenerate SPP modes for the asymmetric trilayers.
Although the non-reciprocity of SPPs should appear for the Voigt configuration even without the hybridization \cite{Hofmann16}, it may vanish due to the mutual interference of SPPs in the symmetric trilayers.
This indicates that difference of the dielectric constants of the two outer insulators is crucial to the non-reciprocity in the trilayer systems.
Chiral magnetic effect in WSMs induces disappearance of the SPP modes, which is similar to that of external static magnetic fields.
For a fixed value of the amplitude of ${\bm g}$, the degree of disappearance of the SPP dispersion curves is smallest for the Faraday configuration, moderate for the perpendicular configuration, and largest for the Voigt configuration.
These results suggest that the Voigt configuration is most suitable to actual applications utilizing the unique properties of WSMs.
Fine-tuning of parameters such as the WSM thickness, dielectric constants, and direction of the axion field enables us to control the interval of the disappearance of SPPs as well as the strength of the non-reciprocity.
We also found the non-monotonic change of the the characteristic decay length for the Faraday configuration as a function of the WSM thickness, which is caused by mode crossing.
These conclusions devise new experiments for identifying the axion filed $\bm{g}$ and exploring the topological nature of WSMs.

Our findings also provide a route for developing new devices. 
We have shown that tuning the dielectric constants of the outer insulators gives rise to stable SRSP modes and causes the LRSP modes to disappear. 
This fact indicates the feasibility of developing plasmonic waveguides, which can reduce the beam radius to nanometer order beyond the diffraction limit of light\cite{Takahara1,Takahara2,Gramotnev}. 
In such a device, the disappearance of LRSP modes would provide an efficient focusing method of the beam radius. 
We also expect that the stable nonreciprocity of SPP dispersion benefits development of a one-way optical waveguide\cite{Takada}, as well as photonic crystals with magneto-optical coupling\cite{Zhu00}. 
These possibilities would provide a way to developing various plasmonic devices exploiting the topological nature of WSMs.

\section*{Acknowledgements}
We acknowledges K. Nomura and H. Ishizuka for useful comments.
This work was supported by JST CREST (JPMJCR14F1) and JST Nanotech CUPAL.

\end{spacing}

\begin{spacing}{1}

%%%%%%%%%% Merge with supplemental materials %%%%%%%%%%
\pagebreak
%\widetext
\begin{center}
\textbf{\large Supplementary materials: Surface plasmon polaritons in thin-film Weyl semimetals}
\end{center}
%%%%%%%%%% Merge with supplemental materials %%%%%%%%%%
%%%%%%%%%% Prefix a "S" to all equations, figures, tables and reset the counter %%%%%%%%%%
\setcounter{equation}{0}
\setcounter{figure}{0}
\setcounter{table}{0}
\setcounter{page}{1}
\makeatletter
\renewcommand{\theequation}{S\arabic{equation}}
\renewcommand{\thefigure}{S\arabic{figure}}
%\renewcommand{\bibnumfmt}[1]{[S#1]}
%\renewcommand{\citenumfont}[1]{S#1}
%%%%%%%%%% Prefix a "S" to all equations, figures, tables and reset the counter %%%%%%%%%%

\section{Maxwell equations in Weyl semimetals}

\begin{figure*}[b]
\begin{center}
\includegraphics[width=5cm]{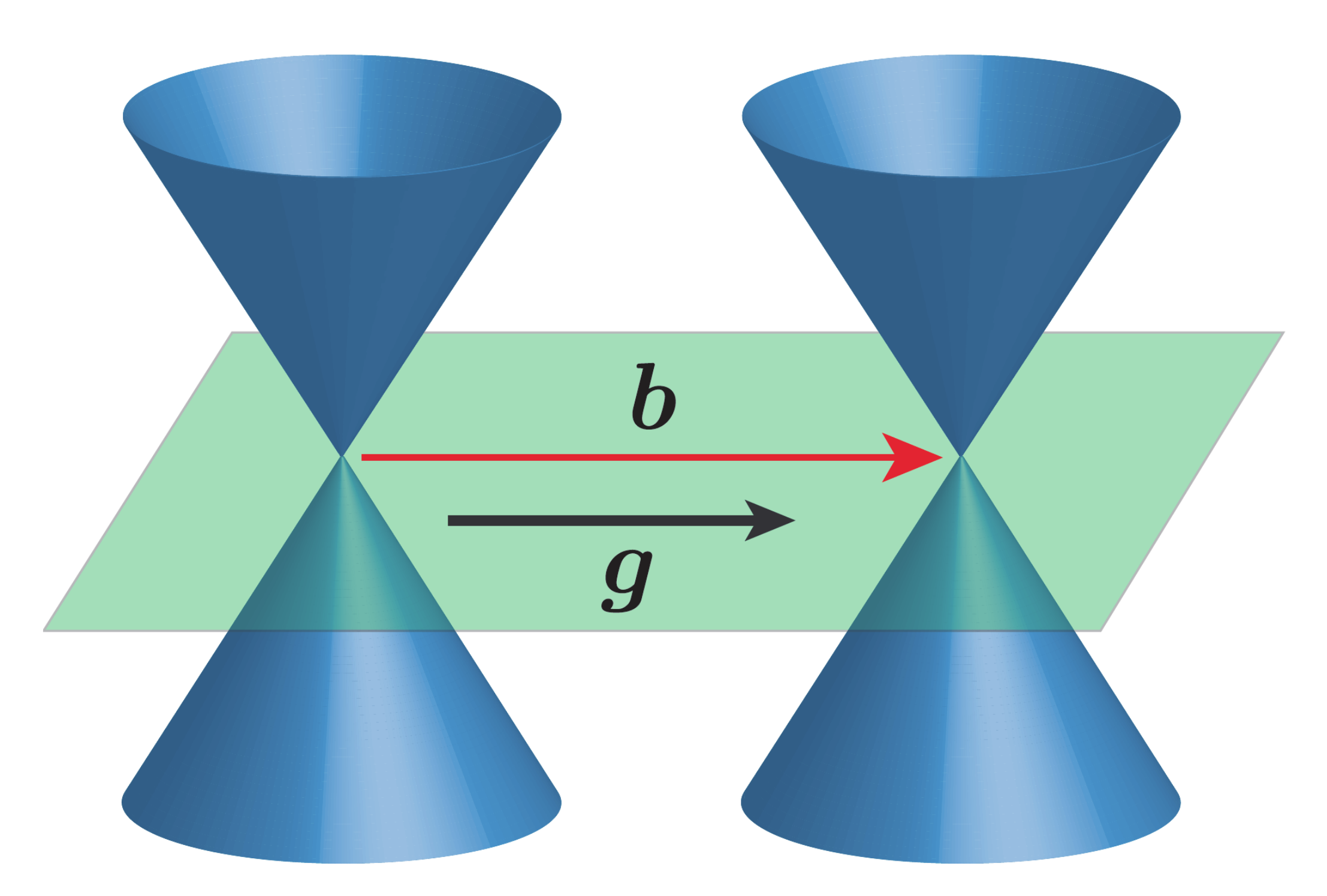}
\caption{(Color online) Schematic diagram of the band structure in WSMs.
Here, ${\mbox{\boldmath $b$}}$ is the distance between Weyl nodes in momentum space.
We define the axion vector $\mbox{\boldmath $g$}$ to be $\bm{g}={2 \alpha_0}/{\pi}  \sqrt{{\epsilon_{0}}/{\mu_{0}}}\bm{b}$, where $\alpha_{0}$, $\epsilon_{0}$,
and $\mu_{0}$ are the fine-structure constant, the dielectric constant, and magnetic permeability, respectively.}
\label{fig:config5}
\end{center}
\end{figure*}
%\linenumbers
In this section, we present a simple derivation and explanation of the Maxwell equations in Weyl semimetals(WSMs).
The optical responses of normal insulators can be expressed by the usual Maxwell equations, whose action is given in the following form: 
\begin{equation}
   S_{0}=\frac{\pi}{8} \int d^{3}x dt\left(\epsilon_{0} {\mbox{\boldmath $E$}^{2}}-{\mbox{\boldmath $B$}^2/\mu_0}\right).
\end{equation}
Here, $\epsilon_{0}$
and $\mu_{0}$ are the dielectric constant and magnetic permeability, respectively, and ${\mbox{\boldmath $E$}}$ and ${\mbox{\boldmath $B$}}$ are the electromagnetic fields inside the insulator.

The unique optical responses in WSMs are derived from the electro-magnetic coupling originating from the topological nature of the materials.
To describe this property, we introduce the additional term in the action of the Maxwell equations represented in the following form\cite{Li,Visinelli}:
\begin{equation}
  S_{\theta}=\frac{\theta}{2\pi}\frac{\alpha_{0}}{2\pi} \int d^{3}x dt {\mbox{\boldmath $E$}} \cdot {\mbox{\boldmath $B$}},
\end{equation}
where $\alpha_{0}=e^{2}/\hbar c$ and $\theta=\theta({\mbox{\boldmath $x$}},t)$ are the fine-structure constant and a pseudo-scalar field known in the particle physics as the axion-like field, respectively.
In the following, we will only consider WSMs with broken time-reversal symmetry , such as $\rm{YbMnB_{2}}$ and $\rm{Eu_{2} Ir_{2} O_{7}}$.
This assumption enables us to simplify the pseudo-scalar field as $\theta({\mbox{\boldmath $x$}},t)=2{\mbox{\boldmath $b$}}\cdot {\mbox{\boldmath $x$}}$, where ${\mbox{\boldmath $b$}}$ is the distance between Weyl nodes in momentum space\cite{Li,Sarma}.

The additional axion term modify the usual Maxwell equations. 
By considering the total action $S_{\rm total}=S_{0}+ S_{\theta}$, we can derive the resulting equations of motion for the ${\mbox{\boldmath $E$}}$ and ${\mbox{\boldmath $B$}}$ fields, with the electric charge $\rho_{e}$ and current densities ${\mbox{\boldmath $J$}_{e}}$ replaced by
\begin{eqnarray}
    & & \rho_{e} \rightarrow  \rho_{e}+\frac{e^{2}}{4 \pi^{2} \hbar} \nabla \theta \cdot {\bf{B}}, \\
    & & {\mbox{\boldmath $J$}_{e}} \rightarrow {\mbox{\boldmath $J$}_{e}}+\frac{e^{2}}{4 \pi^{2} \hbar} \nabla \theta \times {\bf{E}}.
\end{eqnarray}
Thus, we can derive Maxwell equations in WSMs in the form,
\begin{eqnarray}
& & \nabla \times \bm{E}=-\frac{\partial \bm{B}}{\partial t}, \label{eq:maxwell1} \\
& & \nabla \cdot \bm{B}=0,  \label{eq:maxwell2} \\
& & \nabla \cdot \bm{D}=\bm{g} \cdot \bm{B}, \label{eq:maxwell3} \\
& & \nabla \times \bm{H}=\frac{\partial \bm{D}}{\partial t}-\bm{g} \times \bm{E}. \label{eq:maxwell4}
\end{eqnarray}
Here, $\bm{D}$, $\bm{E}$, $\bm{H}$, and $\bm{B}$ are the dielectric displacement, electric field, magnetic field, and magnetic flux density, respectively, and the axion vector ${\mbox{\boldmath $g$}}$ is related to ${\mbox{\boldmath $b$}}$ as 
\begin{equation}
\bm{g}=\frac{2 \alpha_0}{\pi}\sqrt{\frac{\epsilon_{0}}{\mu_{0}}}\bm{b}.
\end{equation}
{In a recent paper~\cite{Kotov18}, the amplitude of $g = |{\bm g}|$ is estimated as about $13\epsilon_0 \omega_P$ for ${\rm Eu}_2{\rm Ir}_2{\rm O}_7$.
Although in the main text we chose a smaller value for $g$ than this estimate, i.e., $g = 4\epsilon_0 \omega_P$, qualitative features of SPPs, i.e., non-reciprocity and disappearance of SPP dispersion,  are expected to be common.
We note that $g$ can be tuned experimentally using, e.g., superlattice structures~\cite{Tominaga}.
}

{WSMs with broken time-reversal symmetry can be realized also in Dirac semimetals such as ${\rm Cd}_3{\rm As}_2$ (Refs.~\cite{Borisenko14,Liang15,Neupane14}), ${\rm ZrTe}_5$ (Ref.~\cite{Li16}), and ${\rm Na}_3{\rm Bi}$ (Ref.~\cite{Liu14}) when an external magnetic field is applied.
WSMs have been reported also for TaAs (Refs.~\cite{Xu99,Lv00}) and for NbAs (Ref.~\cite{Xu999}).
These two materials have time-reversal symmetries, while they have no inversion symmetry.
Although the SPPs in the WSMs with broken inversion symmetry has a unique property as discussed in Ref.~\cite{Sarma}, we do not treat them in this paper to simplify the discussion.}

\section{Detailed information on calculation}
\label{app:deteq}

\subsection{Perpendicular configuration}
\label{app:perpendicular}
As written in the main text, the solution of the modified Maxwell equation for $H_{y}(z)$ in the perpendicular configuration can be expressed as
\begin{eqnarray}
H_y(z) = H_y^{++} e^{-\kappa_+ z} + H_y^{+-} e^{-\kappa_- z} 
+ H_y^{-+} e^{\kappa_+ z} + H_y^{--} e^{\kappa_- z},
\end{eqnarray}
for $|z|<a$, where $H_y^{++}$, $H_y^{+-}$, $H_y^{-+}$, and $H_y^{--}$ are determined by the linear equations,
\begin{eqnarray}
\hat{U}(q,\omega) \left( \begin{array}{c} 
H_{y}^{++} \\ H_{y}^{+-} \\ H_{y}^{-+}  \\ H_{y}^{--}
\end{array} \right) = 0.
\label{eq:defU}
\end{eqnarray}
The $\hat{U}(q,\omega)$ are $4 \times 4$ matrices whose elements are as follows:
\begin{eqnarray}
     & & U_{11} =  e^{ \kappa_+ a} \kappa_- [-\epsilon(\omega)\kappa_+ \kappa_1 - \epsilon_1 \bar{\kappa}^2], \\
     & & U_{12} =  e^{ \kappa_- a} \kappa_+ [-\epsilon(\omega)\kappa_- \kappa_1 - \epsilon_1 \bar{\kappa}^2], \\
     & & U_{13} = -e^{-\kappa_+ a} \kappa_- [ \epsilon(\omega)\kappa_+ \kappa_1 - \epsilon_1 \bar{\kappa}^2], \\
     & & U_{14} = -e^{-\kappa_- a} \kappa_+ [ \epsilon(\omega)\kappa_- \kappa_1 - \epsilon_1 \bar{\kappa}^2], \\
     & & U_{21} =  e^{ \kappa_+ a} \kappa_- (\kappa_+ + \kappa_1), \\
     & & U_{22} = -e^{ \kappa_- a} \kappa_+ (\kappa_- + \kappa_1), \\
     & & U_{23} =  e^{-\kappa_+ a} \kappa_- (\kappa_+ - \kappa_1), \\
     & & U_{24} = -e^{-\kappa_- a} \kappa_+ (\kappa_- - \kappa_1), \\
     & & U_{31} =  e^{-\kappa_+ a} \kappa_- [ \epsilon(\omega)\kappa_+ \kappa_2 - \epsilon_2 \bar{\kappa}^2 ], \\
     & & U_{32} =  e^{-\kappa_- a} \kappa_+ [ \epsilon(\omega)\kappa_- \kappa_2 - \epsilon_2 \bar{\kappa}^2], \\
     & & U_{33} =  e^{ \kappa_+ a} \kappa_- [ \epsilon(\omega)\kappa_+ \kappa_2 + \epsilon_2 \bar{\kappa}^2], \\
     & & U_{34} =  e^{ \kappa_- a} \kappa_+ [ \epsilon(\omega)\kappa_- \kappa_2 + \epsilon_2 \bar{\kappa}^2], \\
     & & U_{41} =  e^{-\kappa_+ a} \kappa_- (\kappa_+ - \kappa_2), \\
     & & U_{42} = -e^{-\kappa_- a} \kappa_+ (\kappa_- - \kappa_2), \\
     & & U_{43} =  e^{ \kappa_+ a} \kappa_- (\kappa_+ + \kappa_2), \\
     & & U_{44} = -e^{ \kappa_- a} \kappa_+ (\kappa_- + \kappa_2), 
\end{eqnarray}
where the definitions of $\kappa_i$ ($i=1,2$) and $\kappa_{\pm}$ are given in the main text, and $\bar{\kappa}^2 \equiv (\kappa_+^2 + \kappa_-^2)/2$. The dispersion is determined by the condition ${\rm det} \, \hat{U}(q,\omega) = 0$. Note that the matrix elements of $\hat{U}$ include the wavenumber $q$ only via $\kappa_{\pm}$. Since $\kappa_{\pm}$ is an even function of $q$, the resultant SPP dispersion has reciprocity, i.e., the relationship $\omega(-q) = \omega(q)$. In the long-wavelength limit ($q\rightarrow 0$), it can be checked that the linear dispersion, $\omega = cq/\sqrt{\epsilon_i}$ ($i=1,2$), is a solution of ${\rm det} \, \hat{U} = 0$, using $\kappa_{\pm} \simeq \sqrt{\epsilon_\infty} \omega_{\rm P}/c$ for $\omega \rightarrow 0$. In the short-wavelength limit ($q\rightarrow \infty$), the SPP frequencies are shown to approach $\Omega_i = \omega_{\rm P}/\sqrt{1+\epsilon_i/\epsilon_\infty}$, using $\kappa \simeq \kappa_1 \simeq \kappa_2 \simeq |q|$ and $e^{-\kappa_{\pm} a}\simeq 0$.

In the limit of $a \to \infty$, $\hat{U}$ becomes block-diagonal, and the linear equation can be rewritten in the form,
\begin{eqnarray}
& & \hat{U}^{(1)}(q,\omega) \left( \begin{array}{c} 
H_{y}^{++} \\ H_{y}^{+-} \end{array} \right) = 0,
\label{eq:defU1} \\
& & \hat{U}^{(2)}(q,\omega) \left( \begin{array}{c} 
H_{y}^{-+}  \\ H_{y}^{--}
\end{array} \right) = 0,
\label{eq:defU2}
\end{eqnarray}
where $\hat{U}^{(1)}$ and $\hat{U}^{(2)}$ are $2\times2$ matrices independent of $a$. Then, the SPP modes in the WSMs are completely decoupled into two independent SPP modes localized at the two interfaces, each of which is characterized by the dielectric constant of either insulator. These indicial equations for $a \to \infty$ coincide with the ones for the two-layer system~\cite{Sarma}.

\subsection{Faraday configuration}
\label{app:Faraday}
For the Faraday configuration, the matrix elements of $\hat{U}(q,\omega)$ are replaced with
\begin{eqnarray}
     & & U_{11} =  \eta_+ e^{ \kappa_+ a}(\kappa_+ + \kappa_1), \\
     & & U_{12} =  \eta_- e^{ \kappa_- a}(\kappa_- + \kappa_1), \\
     & & U_{13} = -\eta_+ e^{-\kappa_+ a}(\kappa_+ - \kappa_1), \\
     & & U_{14} = -\eta_- e^{-\kappa_- a}(\kappa_- - \kappa_1), \\
	 & & U_{21} =         e^{ \kappa_+ a}(\epsilon_1 \kappa_+ + \epsilon(\omega)\kappa_1), \\
     & & U_{22} =         e^{ \kappa_- a}(\epsilon_1 \kappa_- + \epsilon(\omega)\kappa_1), \\
     & & U_{23} = -       e^{-\kappa_+ a}(\epsilon_1 \kappa_+ - \epsilon(\omega)\kappa_1), \\
     & & U_{24} = -       e^{-\kappa_- a}(\epsilon_1 \kappa_- - \epsilon(\omega)\kappa_1), \\
     & & U_{31} =  \eta_+ e^{-\kappa_+ a}(\kappa_+ - \kappa_2), \\
     & & U_{32} =  \eta_- e^{-\kappa_- a}(\kappa_- - \kappa_2), \\
     & & U_{33} = -\eta_+ e^{ \kappa_+ a}(\kappa_+ + \kappa_2), \\
     & & U_{34} = -\eta_- e^{ \kappa_- a}(\kappa_- + \kappa_2), \\
     & & U_{41} =         e^{-\kappa_+ a}(\epsilon_2 \kappa_+ -\epsilon(\omega)\kappa_2), \\
     & & U_{42} =         e^{-\kappa_- a}(\epsilon_2 \kappa_- -\epsilon(\omega)\kappa_2), \\
     & & U_{43} = -       e^{ \kappa_+ a}(\epsilon_2 \kappa_+ +\epsilon(\omega)\kappa_2), \\
     & & U_{44} = -       e^{ \kappa_- a}(\epsilon_2 \kappa_- +\epsilon(\omega)\kappa_2), 
\end{eqnarray}
where $\eta_{\pm}=(\omega^2 \mu_{0} \epsilon_{0}\epsilon(\omega)+\kappa_{\pm}^2-q^2)/g q$ and $\epsilon(\omega)=\epsilon_\infty (1-\omega_{P}^{2}/\omega^{2})$, and $\omega_{P}$ is the plasma frequency of the WSMs. In a similar way as the perpendicular configuration, we can show the following properties of the SPP dispersion: (a) the reciprocity $\omega(-q) = \omega(q)$, (b) the dispersion $\omega = cq/\sqrt{\epsilon_i}$ in the long-wavelength limit ($q\rightarrow 0$), (c) the frequencies $\Omega_i = \omega_{\rm P}/\sqrt{1+\epsilon_i/\epsilon_\infty}$ in the short-wavelength limit ($q\rightarrow \infty$), and (d) the decoupled SPPs described by the linear equations, Eqs.~(\ref{eq:defU1}) and (\ref{eq:defU2}), for $a \rightarrow \infty$.

\subsection{Voigt configuration}
\label{app:Voigt}
For the Voigt configuration, the TE and TM modes are not mixed by the axion field. The profile of the transverse magnetic field in the WSMs is given as
\begin{eqnarray}
H_y(z) = H_y^+ e^{-\kappa z} + H_y^- e^{\kappa z}, 
\end{eqnarray}
for $|z|<a$, where the constants, $H_y^{+}$ and $H_y^{-}$, are determined by the linear equations,
\begin{eqnarray}
\hat{V} \left( \begin{array}{c} 
H_y^+ \\ H_y^- 
\end{array} \right) = 0.
\label{eq:defV}
\end{eqnarray}
The matrix elements of $\hat{V}$ are 
\begin{eqnarray}
    & & V_{11} = e^{\kappa a}\left[\frac{\omega \epsilon_{0} \epsilon(\omega) \kappa - qg}{(\omega \epsilon_{0} \epsilon(\omega))^2-g^2}
    +\frac{\kappa_1}{\omega \epsilon_{0} \epsilon_1}\right], \\
    & & V_{12} = e^{-\kappa a}\left[-\frac{\omega \epsilon_{0} \epsilon(\omega) \kappa + qg}{(\omega \epsilon_{0} \epsilon(\omega))^2-g^2}
    +\frac{\kappa_1}{\omega \epsilon_{0} \epsilon_1}\right], \\
    & & V_{21} = e^{-\kappa a}\left[-\frac{\omega \epsilon_{0} \epsilon(\omega) \kappa - qg}{(\omega \epsilon_{0} \epsilon(\omega))^2-g^2}
    +\frac{\kappa_2}{\omega \epsilon_{0} \epsilon_2}\right], \\
    & & V_{22} = e^{\kappa a}\left[\frac{\omega \epsilon_{0} \epsilon(\omega) \kappa + qg}{(\omega \epsilon_{0} \epsilon(\omega))^2-g^2}
    +\frac{\kappa_2}{\omega \epsilon_{0} \epsilon_2}\right], 
\end{eqnarray}
Nontrivial solutions of (\ref{eq:defV}) exist when the condition ${\rm det} \, \hat{V}(q,\omega)= 0$ holds. This condition can be rewritten as
\begin{eqnarray}
    & & e^{-4\kappa a} = \frac{\alpha_+ \beta_+}{\alpha_- \beta_-},
    \label{eq:VoigtSPP1} \\
    & & \alpha_{\pm} = \kappa_1((\omega \epsilon_{0} \epsilon(\omega))^2-g^2)\pm\epsilon_{0}\epsilon_1\omega (\epsilon_{0} \epsilon(\omega)\omega \kappa - q g), 
        \label{eq:VoigtSPP2} \\
    & & \beta_{\pm}  = \kappa_2((\omega \epsilon_{0} \epsilon(\omega))^2-g^2)\pm\epsilon_{0}\epsilon_2\omega (\epsilon_{0} \epsilon(\omega)\omega \kappa + q g),
        \label{eq:VoigtSPP3} 
\end{eqnarray}
and determines the SPP dispersion $\omega = \omega(q)$.

For the Voigt configuration, the SPP dispersion has non-reciprocity, since the determination equation for the SPP dispersion is not invariant under reversal of the propagation direction ($q \rightarrow -q$). The non-reciprocity for the SRSP mode can clearly be seen in the limit $q \rightarrow \pm \infty$, for which $\kappa \simeq \kappa_1 \simeq \kappa_2 \simeq |q|$ and $e^{-4\kappa a}\simeq 0$. For $\epsilon_2 > \epsilon_1$, the frequencies of SRSP are calculated from Eqs.~(\ref{eq:VoigtSPP1})-(\ref{eq:VoigtSPP3}) as follows:
\begin{eqnarray}
& & \Omega_1 \equiv \omega(q\rightarrow + \infty) 
= \frac{- g+\sqrt{g^2+4\epsilon_\infty \epsilon^2_0 \omega^2_p(\epsilon_\infty+\epsilon_1)}}{2\epsilon_{0} (\epsilon_\infty+\epsilon_1)}, \label{eq:Omega1} \\
& & \Omega_2 \equiv \omega(q\rightarrow - \infty) 
= \frac{- g+\sqrt{g^2+4\epsilon_\infty \epsilon^2_0\omega^2_p(\epsilon_\infty+\epsilon_2)}}{2\epsilon_{0} (\epsilon_\infty+\epsilon_2)},
\end{eqnarray}
We should note that the LRSP dispersion disappears in the high-frequency region because the condition $\kappa^2 > 0$ is not satisfied. These expressions indicate non-reciprocity, i.e., {$\Omega_1 \ne \Omega_2$} when ${\bm g}\ne {\bm 0}$.

In the limit $a \to \infty$, the matrix $\hat{V}$ becomes diagonal, and the two SPP modes are reduced to two independent SPP modes localized at the two interfaces.
The dispersions of these two independent SPP modes are determined by $V_{11} = 0$ and $V_{22} = 0$, respectively.

\section{Faraday configuration in two-layer system}
\label{app:FaradayTwoLayer}

\begin{figure*}[tbh]
\begin{center}
\includegraphics[width=14cm]{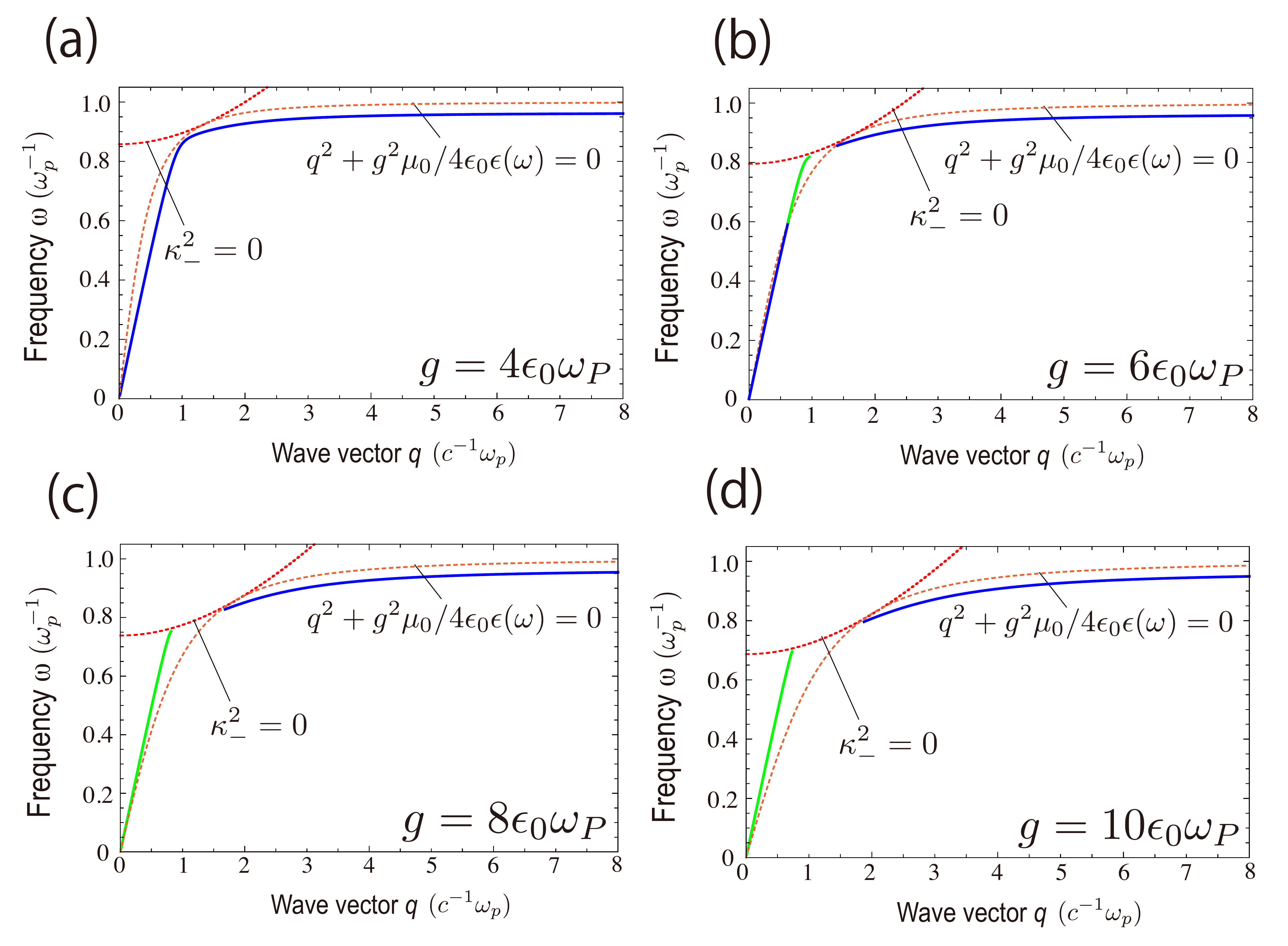}
\caption{(Color online) The SPP dispersion curves for the Faraday configuration; (a) $g=4\epsilon_{0} \omega_p$, (b) $6\epsilon_{0} \omega_p$, (c) $8\epsilon_{0} \omega_p$, and (d) $10\epsilon_{0} \omega_p$. Here, the red and orange dotted lines indicate the conditions of $\kappa_{-}=0$ and $K=0$, respectively. The green and blue lines show the SPP dispersions with non-oscillating and oscillating decay modes, respectively.}
\label{fig:config3}
\end{center}
\end{figure*}

\begin{figure*}[tbh]
\begin{center}
\includegraphics[width=14cm]{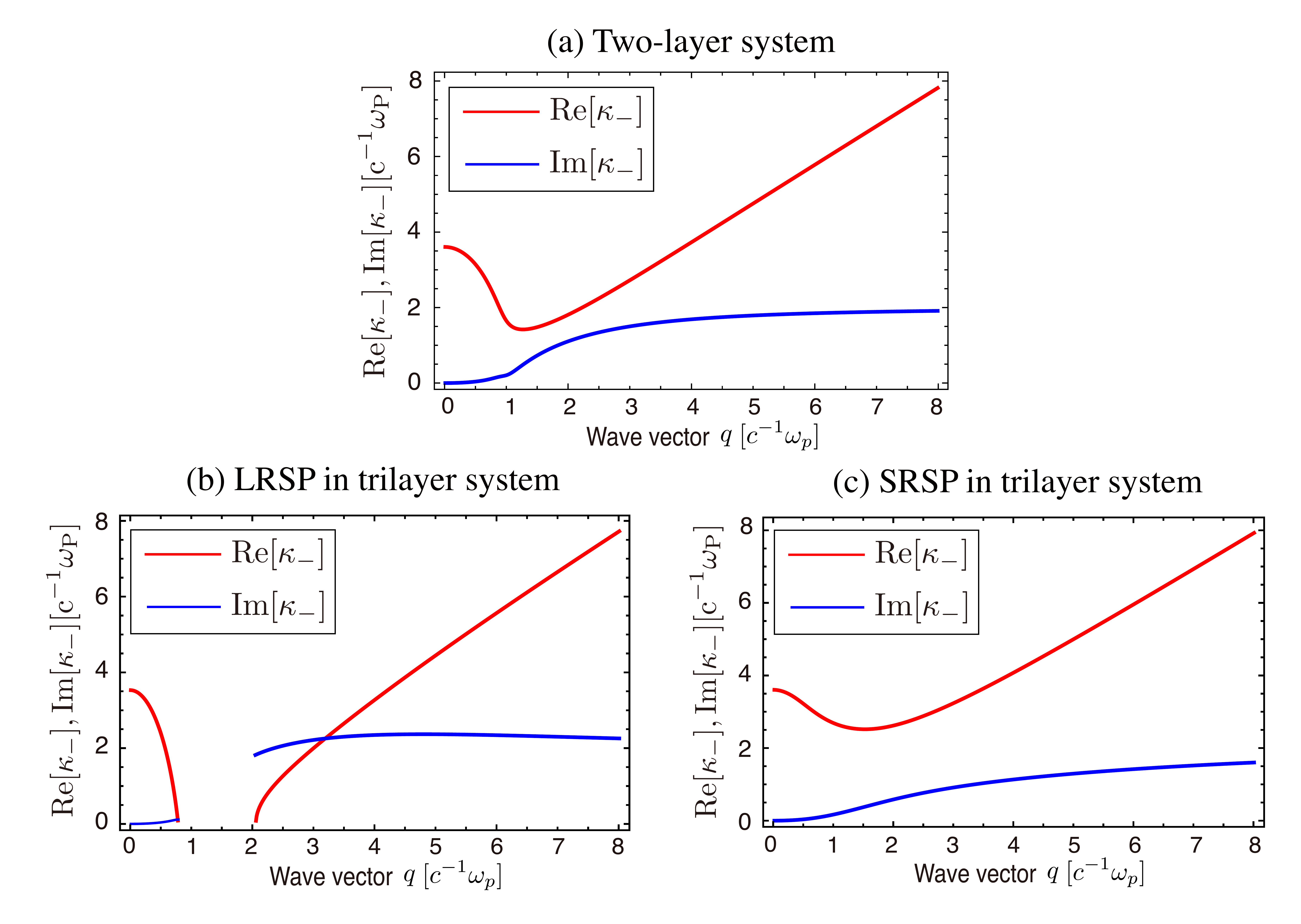}
\caption{(Color online) Real and imaginary parts of $\kappa_{-}$ for the Faraday configuration; (a) the bilayer system and (b) the LRSP and (b) the SRSP for the symmetric trilayer system.
The red and blue lines indicate the real and imaginary parts of $\kappa_{-}$ for $2a=0.2c \omega_{P}^{-1}$ and $g=4\epsilon_{0} \omega_p$ . 
Fig. (b) and (c) show the values of $\kappa_{-}$ for LRSP and SRSP modes, respectively.} 
\label{fig:config4}
\end{center}
\end{figure*}

In this section, we discuss the ${\bm{g}}$-dependence of SPPs in the bilayer system for the Faraday configuration. Figure~\ref{fig:config3}~ plots the SPP dispersion for (a) $g=4\epsilon_{0} \omega_p$, (b) $6\epsilon_{0} \omega_p$, (c) $8\epsilon_{0} \omega_p$, and (d) $10\epsilon_{0} \omega_p$. Here, the red and orange dotted lines show the conditions of $\kappa_{-}=0$ and $K=0$, where
\begin{eqnarray}
& & {\kappa}^{2}_{-}=q^2-\epsilon(\omega) \omega^2/c^2 +\frac{g^2\mu_{0}}{2\epsilon_0 \epsilon(\omega)} \pm g \sqrt{K}, \\
& & K = \frac{\mu_{0}}{\epsilon_0 \epsilon(\omega)} \left(q^2+ \frac{g^2 \mu_0}{4\epsilon_{0}\epsilon(\omega)} \right),
\end{eqnarray}
The blue and green lines in Fig.~\ref{fig:config3} indicate the SPP dispersion for complex solutions (complex ${\kappa}_{\pm}$) and real solutions (real ${\kappa}_{\pm}$), respectively. Here, we set $\epsilon_{\infty}=13$ and $\epsilon_{1}=\epsilon_{2}=1$. Figure~\ref{fig:config3}~(a) shows that $g=4\epsilon_{0} \omega_p$ only yields complex SPP modes, whose wavefunctions show oscillating decay in the stacking direction. 
With increasing $g$, the complex solutions change into the real ones around $q=0.75c^{-1}\omega_p$ as can be seen in Fig. ~\ref{fig:config3}~(b). 
Around $q=0$, the complex solutions disappear at $g=2\sqrt{\epsilon_{1} \epsilon_{2}}\epsilon_{0} \omega_p$. 
We can conclude that SPP dispersion with both non-oscillating and oscillating decay modes emerge in the Faraday configuration.

Figure~\ref{fig:config4} shows the real and imaginary part of ${\kappa}_{-}$ in the case of $g=4\epsilon_{0} \omega_p$ for (a) the bilayer system and (b-c) the symmetric trilayer system, respectively. 
Here, the red and blue lines indicate ${\rm{Re}}[{\kappa}_{-}]$ and ${\rm{Im}}[{\kappa}_{-}]$.
Figure ~\ref{fig:config4}(a) shows that there exists the distorted point around $q=1.0c^{-1}\omega_p \sim 1.5c^{-1}\omega_p$, above which the variation of ${\rm{Re}}[{\kappa}_{-}]$ as a function of $q$ significantly changes.
By comparing to Figure~\ref{fig:config3}(a), we identify that this point corresponds to the most nearest area to the threshold of $\kappa_{-}^{2}=0$.
This figure indicates that the characteristic decay length in Faraday configuration has maximum values as a function of $q$.
Figures~\ref{fig:config4}(b) and (c) show $\kappa_{-}$ for the LRSP and SRSP modes, respectively.
In this numerical calculation, we suppose the thickness of WSMs to be $2a=0.2c \omega_{P}^{-1}$.
Figure~\ref{fig:config4}(b) shows the disappearance in ${\rm{Re}}[{\kappa}_{-}]$ and ${\rm{Im}}[{\kappa}_{-}]$ at around $q=1.0c^{-1}\omega_p \sim 2.0c^{-1}\omega_p$.
In this region, ${\rm{Re}}[{\kappa}_{-}]$ indicates zero values, and consequently, show the bulk-propagating modes corresponding to the vanishing area in LRSP modes.
By combining the result in Figure~\ref{fig:config4}(a), we can conclude that the LRSP (SRSP) mode becomes unstable (stable), respectively, comparing to that of the bilayer system.

\section{Characteristic decay length of $\kappa_{+}$ in the perpendicular configuration}
\label{app:WeakDependence}

\begin{figure*}[b!]
\begin{center}
\includegraphics[width=14cm]{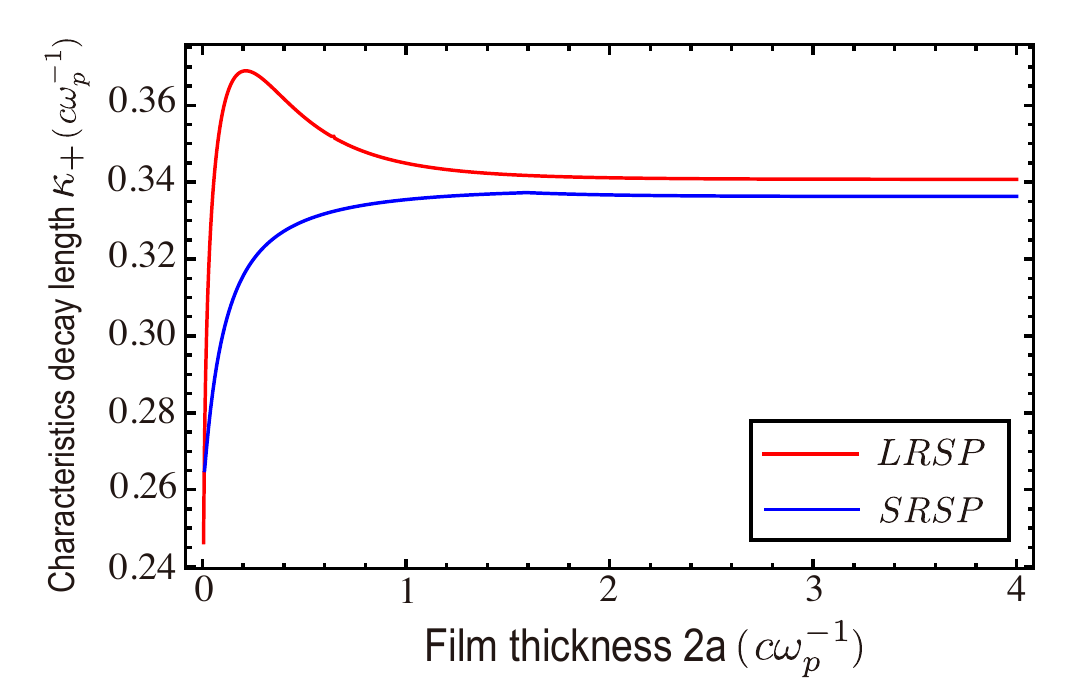}
\caption{(Color online) Characteristic decay length of $\kappa_{+}$ as a function of WSM thickness for the perpendicular configuration. The dielectric constants of the insulators are $\epsilon_1=\epsilon_2=1$. The wavenumber of SPPs is $q=1.0c^{-1}\omega_p$.}
\label{fig:config5}
\end{center}
\end{figure*}

We show in figure \ref{fig:config5} the characteristic decay length of $\kappa_{+}$ as a function of WSM thickness for the perpendicular configuration. 
Here, we assume the dielectric constants of the insulators and the wavenumber of SPPs to be $\epsilon_1=\epsilon_2=1$ and $q=1.0c^{-1}\omega_p$, respectively.
This figure shows that the LRSP and SRSP modes indicate small variations, $0.24 c \omega_{p}^{-1}\leq \kappa_{+}^{-1} \le 0.37c \omega_{p}^{-1}$, compared to that of the $\kappa_{-}$ in the main-text Figure.5.
Based on this data, we only focus the characteristic decay length of $\kappa_{-}$ and discuss their properties in the main text.

\end{spacing}

\begin{spacing}{1}

\end{spacing}
\end{document}